\documentclass[aps,prd,preprint,groupedaddress]{revtex4}
\usepackage{epsfig}

\begin{document}

\def\Re  {{\rm Re}}
\def\Im  {{\rm Im}}
\def\KFS#1{K^{{#1}}_{\scriptscriptstyle\rm F\!.S\!.}}
\def\Ret {\widetilde{{\rm Re}}}
\def\veps{\varepsilon}
\def\hc    {{\rm h.c.}}
\newcommand{\fact}{{\mathrm{C}}}
\newcommand{\figref}[1]{Fig.~\ref{#1}}
\newcommand{\tabref}[1]{Tab.~\ref{#1}}
\newcommand{\rA}{{\mathrm{A}}}
\newcommand{\rB}{{\mathrm{B}}}
\newcommand{\rR}{{\mathrm{R}}}
\newcommand{\rV}{{\mathrm{V}}}
\def\bom#1{{\mbox{\boldmath $#1$}}}
\def\ket#1{|{#1}\rangle}
\def\bra#1{\langle{#1}|}
\def\sd{\tilde{d}}
\def\sq{\tilde{q}}
\def\su{\tilde{u}}
\def\mq{m_q}
\def\mqi{m_{q}}
\def\mqj{m_{q}}
\def\mqk{m_{q}}
\def\mql{m_{q}}
\def\msqi{m_{\tilde{q}_i}}
\def\msqj{m_{\tilde{q}_j}}
\def\msqk{m_{\tilde{q}_k}}
\def\msql{m_{\tilde{q}_l}}
\def\mgl{m_{\tilde{g}}}

\def\as{\alpha_s}
\def\nbar{\bar{N}}
\def\bbar{\bar{b}}

\def\ca{\tilde{\chi}^\pm_1}
\def\cb{\tilde{\chi}^\pm_2}
\def\na{\tilde{\chi}^0_1}
\def\nb{\tilde{\chi}^0_2}
\def\nc{\tilde{\chi}^0_3}
\def\nd{\tilde{\chi}^0_4}
\def\ncd{\tilde{\chi}^0_{3,4}}
\def\sq{\tilde{q}}

\def\cA{{\cal A}} \def\cB{{\cal B}} \def\cC{{\cal C}} \def\cD{{\cal D}}
\def\cE{{\cal E}} \def\cF{{\cal F}} \def\cG{{\cal G}} \def\cH{{\cal H}}
\def\cI{{\cal I}} \def\cJ{{\cal J}} \def\cK{{\cal K}} \def\cL{{\cal L}}
\def\cM{{\cal M}} \def\cN{{\cal N}} \def\cO{{\cal O}} \def\cP{{\cal P}}
\def\cQ{{\cal Q}} \def\cR{{\cal R}} \def\cS{{\cal S}} \def\cT{{\cal T}}
\def\cU{{\cal U}} \def\cV{{\cal V}} \def\cW{{\cal W}} \def\cX{{\cal X}}
\def\cY{{\cal Y}} \def\cZ{{\cal Z}} 

\def\d{{\rm d}}
\def\eps{\epsilon}
\def\MSbar{\overline{\rm MS}}

\def\lp{\left. }
\def\rp{\right. }
\def\lr{\left( }
\def\rr{\right) }
\def\le{\left[ }
\def\re{\right] }
\def\lg{\left\{ }
\def\rg{\right\} }
\def\lb{\left| }
\def\rb{\right| }

\def\bsp#1\esp{\begin{split}#1\end{split}}

\def\beq{\begin{equation}}
\def\eeq{\end{equation}}
\def\bea{\begin{eqnarray}}
\def\eea{\end{eqnarray}}

\preprint{DESY 11-043}
\preprint{MS-TP-11-06}
\title{Dijet photoproduction of massless charm jets \\ at next-to-leading order
 of QCD}
\author{Michael Klasen$^a$}
\email[]{klasen@lpsc.in2p3.fr}
\author{Gustav Kramer$^b$}
\affiliation{$^a$ Institut f\"ur Theoretische Physik, Westf\"alische
 Wilhelms-Universit\"at M\"unster, Wilhelm-Klemm-Stra\ss{}e 9, D-48149 M\"unster,
 Germany\\
 $^b$ II.\ Institut f\"ur Theoretische Physik, Universit\"at Hamburg, Luruper
 Chaussee 149, D-22761 Hamburg, Germany}
\date{\today}
\begin{abstract}
We compute the charm dijet photoproduction cross section at next-to-leading order
of QCD in the zero-mass variable flavour number scheme, i.e.\ with active charm
quarks in the proton and photon. The results are compared to recent measurements
from the ZEUS experiment at HERA. The predictions for various distributions
agree well with the data, in particular for large momentum fractions of the
the partons in the photon, where direct photon processes dominate. At low momentum
fractions, the predictions are quite sensitive to the charm content in the photon.
The experimental data are shown to favour parameterizations with a substantial
charm quark density such as the one proposed by Cornet et al.
\end{abstract}
\maketitle



\section{Introduction}
\label{sec:1}

In photoproduction processes at HERA, a quasi-real photon emitted from the 
incoming electron (or positron) collides with a parton from the incoming 
proton. Within QCD the photoproduction is classified into two categories: 
(i) direct processes, in which the photon acts as a point-like particle in the 
hard scattering, and (ii) resolved processes, in which the photon acts as a 
source of incoming partons which participate in the hard interaction. The two 
classes result directly from the next-to-leading order (NLO) 
QCD calculation of the photoproduction cross sections, 
since due to the initial photon's virtuality $Q^2=0$,
the direct processes in NLO have an initial-state singularity which must be 
absorbed into the photon parton distribution functions (PDFs), on which the 
resolved cross section depends. 

Measurements of cross sections for the production of jets with high transverse 
energy are sensitive to the PDFs of both the proton and the photon. The proton
PDFs are known from the extensive global analyses of the CTEQ \cite{1} and 
MSTW \cite{2} groups. The photon PDFs are essentially based only on data for 
$F_2^{\gamma}(x)$. Of course the proton PDFs are much better known than the 
photon PDFs. The three newest photon PDFs, Cornet et al. (CJK) \cite{3}, 
Aurenche et al. (AFG04) \cite{4} and Slominski et al. (SAL) \cite{5}, use all 
available data of $F_2^{\gamma}$ from the LEP experiments. The older 
parameterizations of Gl\"uck et al. (GRV) \cite{6} and Aurenche et al. (AFG) 
\cite{7} could use only the lower energy measurements of $F_2^{\gamma}$ known 
at the time of their construction and are, of course, less reliable than the 
more recent sets.

In 2007 the ZEUS collaboration at HERA presented their data on high-$E_T$ dijet
photoproduction \cite{8} and compared them to NLO QCD calculations with the 
aim to provide constraints on the PDFs of the photon. The comparison was done 
to the five photon PDFs mentioned above. The differences obtained with these 
PDFs were generally less than $25\%$ for the AFG, AFG04, SAL and GRV sets. The 
predictions based on CJK set was up to $70\%$ higher than those based on the 
other four. These differences occurred predominantly in the small $x_{\gamma}$ 
region and for low $E_T$. Otherwise the agreement with all five PDFs was very 
satisfactory. The photon PDFs have three essentially different components: 
light quarks, heavy quarks (charm and/or bottom) and the gluon. The best 
known of these three components is that of the light quarks, while much less is
known about the charm and bottom quark contents. However, it is possible to 
separate the heavy quarks, for example the charm quark PDF in the photon, from 
the light quark ones by measuring the dijet production cross sections with at 
least one charm quark jet. Such measurements have been done in the past by the
ZEUS collaboration at HERA \cite{9,10,11}. In these measurements the selection of 
the charm jet occurs by measuring the differential cross section for the 
production of a $D^{*}$-meson associated with the dijet system. So far, these 
data have been compared only with NLO predictions based on the massive charm 
scheme or fixed-flavour number scheme (FFNS) \cite{12}, in which an explicit 
charm component of the photon PDFs in the resolved cross section does not occur.
In this scheme such a contribution is approximated with the NLO partonic cross
section, convolved with the PDFs of 
incoming light quarks and the gluon. On the other hand in the
comparison of the inclusive dijet photoproduction data (see for example 
\cite{8} and the earlier work quoted there) with the NLO calculations, the 
charm quark is always considered massless with the consequence that in the 
resolved contribution the charm component of the photon PDFs is present. This 
approach would be justified when the cross section with at least one charm 
could satisfactorily be described in the massless charm scheme. 
 
It is the purpose of this work to present results of NLO calculations in the
massless charm scheme for dijet cross sections with at least one charm jet and 
to find out whether they can describe satisfactorily the measured cross 
sections from ZEUS \cite{10}. There, a similar comparison had been performed,
which was, however, based only on leading order Monte Carlo simulations.
Of course such a comparison depends on the charm
PDF of the photon. The treatment of the charm quark being massless is 
justified as long as the $E_T$ of the produced jets is large enough, i.e.
$E_T^2 \gg m^2$, where $m$ is the charm quark mass. The ZEUS data reported in  
\cite{10} are particularly suitable for our purpose, since by measuring the 
$x_{\gamma}$ distribution $d\sigma/dx_{\gamma}$ and the angular distribution
of the outgoing jets for large and small $x_{\gamma}$, i.e. for the dominant 
subprocesses, direct and resolved, the data are most sensitively dependent 
on the charm content of the photon PDF.
 
In section 2 we shall describe the theoretical framework and outline the 
kinematical restrictions based on the experimental cuts applied in \cite{10}. 
Section 3 contains our results and the discussion of the comparison with the 
data \cite{10} for the default choice of the GRV photon PDF. In section 4 we 
compare different photon PDFs for a scale characteristic for charm jet
production, and in section 5 we give cross sections obtained with three
different photon PDFs, GRV, AFG04 and CJK. Our conclusions are presented in 
section 6.

\section{Theoretical Framework and Kinematical Constraints} 

For our calculations we rely on our work on dijet production in the reaction 
$\gamma + p \to jets + X$ \cite{13} in which we have calculated the cross 
section for inclusive one-jet and two-jet production up to NLO for both the 
direct and the resolved contribution (for a review see \cite{14}). The 
predictions of this work have been tested by many experimental studies of the 
H1 and ZEUS collaborations at HERA. To obtain the cross section with at least 
one charm jet in the final state we calculated the difference of the 
cross section with $n_f=4$ quark flavours and the cross section with $n_f=3$ 
quark flavours in the initial state. In the resolved part this difference 
contains only the contribution of $cg,\bar{c}g,cq,\bar{c}q$ and $c\bar{c}$ in 
the initial state, where $q$ is a light quark (or antiquark). In the $c\bar{c}$
contribution it contains also the contribution from $c\bar{c} \to gg$ and 
$c\bar{c} \to q\bar{q}$ in leading order (LO) and the corresponding 
contributions in NLO. Since these contributions originate from the charm PDFs 
of the photon and the proton, their magnitude is expected to be small. In LO 
their contribution to the cross section integrated over the whole phase region 
as in the ZEUS experiment, to be specified below, amounts to $0.3\%$. We expect
a similar amount from the NLO contributions. As input we employ the 
NLO GRV 92 photon PDFs \cite{6}, 
converted to the $\overline{MS}$ scheme, and the NLO CTEQ6M proton PDFs 
\cite{15}. The strong-coupling constant $\alpha_s^{(n_f)}(\mu_R)$ is evaluated 
from the two-loop formula \cite{16} with $n_f$ active quark flavours. Both in 
the $n_f=3$ calculation and in the difference of the $n_f=4$ and $n_f=3$ cross 
sections, $n_f=4$ is used in the $\alpha_s$ formula. The asymptotic scale 
parameter is $\Lambda_{\overline{MS}}^{(4)} =0.347$ GeV corresponding to 
$\alpha_s^{(5)}(m_Z) = 0.118$. We choose the renormalization scale $\mu_R$ and 
the factorization scale $\mu_F$ of the initial state to be 
$\mu_R=\xi_R \bar{E}_T$ and $\mu_F=\xi_F \bar{E}_T$, where $\bar{E}_T$ is the
average transverse energy of the two (or three)final-state partons constituting
the two jets. $\xi_R$ and $\xi_F$ are dimensionless scale factors, which are 
varied about their default values $\xi_R=\xi_F=1$ as described later. 
The experimental data presented in  \cite{10} were taken with electrons or 
positrons of energy $E_e=27.5$ GeV and protons with energy $E_p=820$ GeV 
(1996-1997) or $E_p=920$ GeV (1998-2000) corresponding to integrated 
luminosities of $38.6\pm0.6$ and $81.9\pm1.8$ $pb^{-1}$ and to centre-of mass 
energies of $\sqrt{s}=300$ GeV and $\sqrt{s}=318$ GeV, respectively. In our 
calculations we took the kinematic conditions of the 1998-2000 period due to 
its larger luminosity.
 
In the experiment, the photon virtuality $Q^2$ was restricted to be below 
$Q^2_{max}=1~GeV^2$. The photon-proton centre-of mass energy was lying in the 
range $130 < W < 280$ GeV. Jets were reconstructed with the $k_T$-cluster 
algorithm in the longitudinally invariant inclusive mode \cite{17}. The events 
were required to have at least two jets with pseudorapidity 
$|\eta^{jet}| < 2.4$ and transverse energy $E_T > 5.0$ GeV. With the two 
energetic jets a $D^{*}$-meson is observed. The reconstructed $D^{*}$ meson is 
required to have $p_T^{D^{*}} > 3$ GeV and pseudorapitity 
$|\eta^{D^{*}}| < 1.5$. 

In \cite{10}, the following three jet variables were chosen in the analysis:
$x_{\gamma}^{obs}$, $x_p^{obs}$ and $\cos\Theta^{*}$. In order to select 
contributions to the cross section enriched by direct and resolved photon 
events, the variable $x_{\gamma}^{obs}$ defined by
\begin{eqnarray}
 x_{\gamma}^{obs}= \frac{\sum_{jets}(E_T^{jet}e^{-\eta^{jet}})}{2yE_e} 
\end{eqnarray}
was determined. Here, $yE_e$ is the initial photon energy, and the sum in 
Eq.\ (1) is over the two jets with the two highest $E_T^{jet}$. Due to the 
restriction to two jets, $x_{\gamma}^{obs}$ does not measure 
the full $x_{\gamma}$, whence
the superscript ``obs''. With $x_{\gamma}^{obs} > (<) 0.75$ samples enriched in
direct (resolved) photon processes are selected. The complementary variable on 
the proton side is
\begin{equation}
  x_p^{obs} =\frac{\sum_{jets}(E_T^{jet} e^{\eta^{jet}})}{2E_p},
\end{equation}
which is the fraction of the proton's momentum contributing to the production 
of the two jets with highest $E_T^{jet}$. The third variable is 
$\cos\Theta^{*}$, where $\Theta^{*}$ is the dijet scattering angle. It is 
given by  
\begin{equation}
  \cos\Theta^{*} = \tanh((\eta^{jet1}-\eta^{jet2})/2).
\end{equation}
In order to enhance the characteristic features of the direct versus the 
resolved contributions in the $\cos\Theta^{*}$ distribution, a cut on the 
invariant mass $M_{jj}$ is applied: $M_{jj}>18$ GeV. For the case in which the 
two jets are back-to-back in the transverse plane with equal transverse $E_T$,
the dijet invariant mass is
\begin{equation}
 M_{jj} =\frac{2E_T^{jet}}{\sqrt{1-|\cos\Theta^*|^2}}.
\end{equation}
In order to achieve high values of $|\cos\Theta^{*}|$ without bias from the 
$E_T^{jet}$ cut, $M_{jj}$ must be large enough. In our calculations we shall 
study only the distribution in the absolute value of $\cos\Theta^{*}$, which 
does not depend on which is jet number 1 and jet number 2. 
In the data analysis of \cite{10} a cut on the average longitudinal boost 
$\bar{\eta} =(\eta^{jet1}+\eta^{jet2})/2$ of $\bar{\eta} < 0.7$ is applied. 
This selection limits $\eta^{jet}$ to $|\eta^{jet}| < 1.9$, removes the bias
caused by the explicit cut on $\eta^{jet}$ and reduces the bias caused by the 
cut on $|\eta^{D^{*}}| < 1.5$. According to \cite{10} the residual distortion 
due to $\eta^{D^{*}}$ cut is small and confined to the extreme bins of the 
$\cos\Theta^{*}$ distribution. With all these cuts it is achieved that the 
measured distributions test to a large extent the dynamics of the hard 
scattering processes.

All these experimented cuts on the rapidities $\eta^{jet}$, transverse energies
$E_T^{jet}$ and the invariant mass $M_{jj}$ are also applied to the calculation
of the cross sections in NLO. However, the $E_T^{jet} \geq 5.0$ GeV cut to both
jets cannot be applied, since it leads to infrared sensitive cross sections 
\cite{18}. Therefore we choose $E_T^{jet1} \geq 5.5$ GeV and 
$E_T^{jet2} \geq 5.0$ GeV. Unfortunately, in the experimental data analysis 
such an asymmetric cut has not been applied, although in an earlier \cite{9} 
and the later publication \cite{11} such asymmetric cuts were used to make the
comparison with the results of the massive NLO calculations possible. Since 
with our choice the difference between the two cut values is small, $0.5$ GeV, 
we expect that it has little influence on the measured cross sections.

The observed $D^{*}$ in every event triggers the production of at least one 
charm jet. The fraction of charm quarks fragmenting into a $D^{*}$ meson was 
assumed to be $0.235$ \cite{19}. We multiply the calculated charm jet cross 
section with this branching ratio when we compare to the experimental data. 
Actually the measured cross sections are the luminosity-weighted average of the
cross sections at the two proton energies $E_p=820$ GeV and $E_p=920$ GeV. In 
our calculation we neglect this averaging and give only results for $E_p=920$
GeV. 

\section{Results for GRV photon PDFs and comparison with ZEUS data}
%
\begin{figure}
 \centering
 \epsfig{file=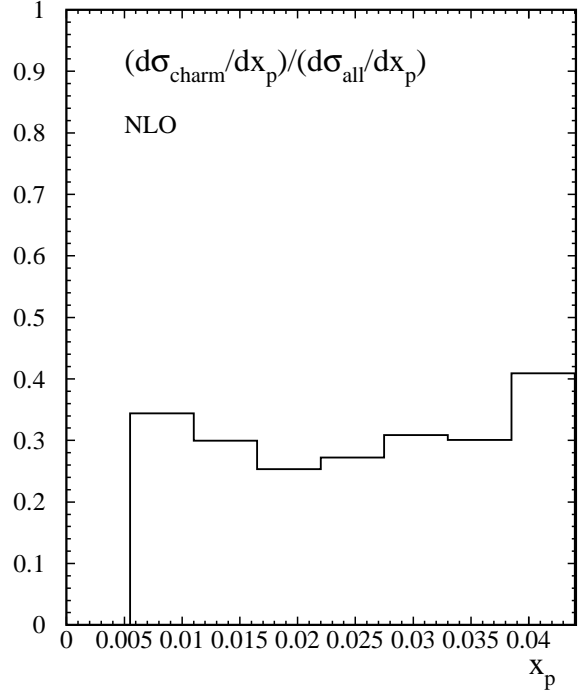,width=.55\columnwidth}
\caption{\label{fig:1} Ratio of the charm dijet cross section $d\sigma/dx_p$
to the full dijet cross section (all flavours up to charm) as a function of 
$x_p$.}
\end{figure}
%
\begin{figure}
 \centering
\epsfig{file=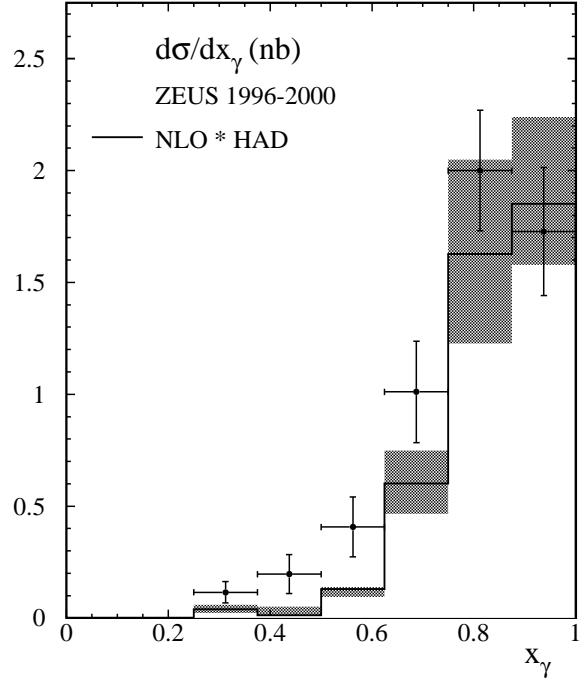,width=.55\columnwidth}
 \caption{\label{fig:2} Differential cross section $d\sigma/dx_{\gamma}$ as a 
function of $x_\gamma$ compared to the data of \cite{10}.}
\end{figure}
%
\begin{figure}
 \centering
\epsfig{file=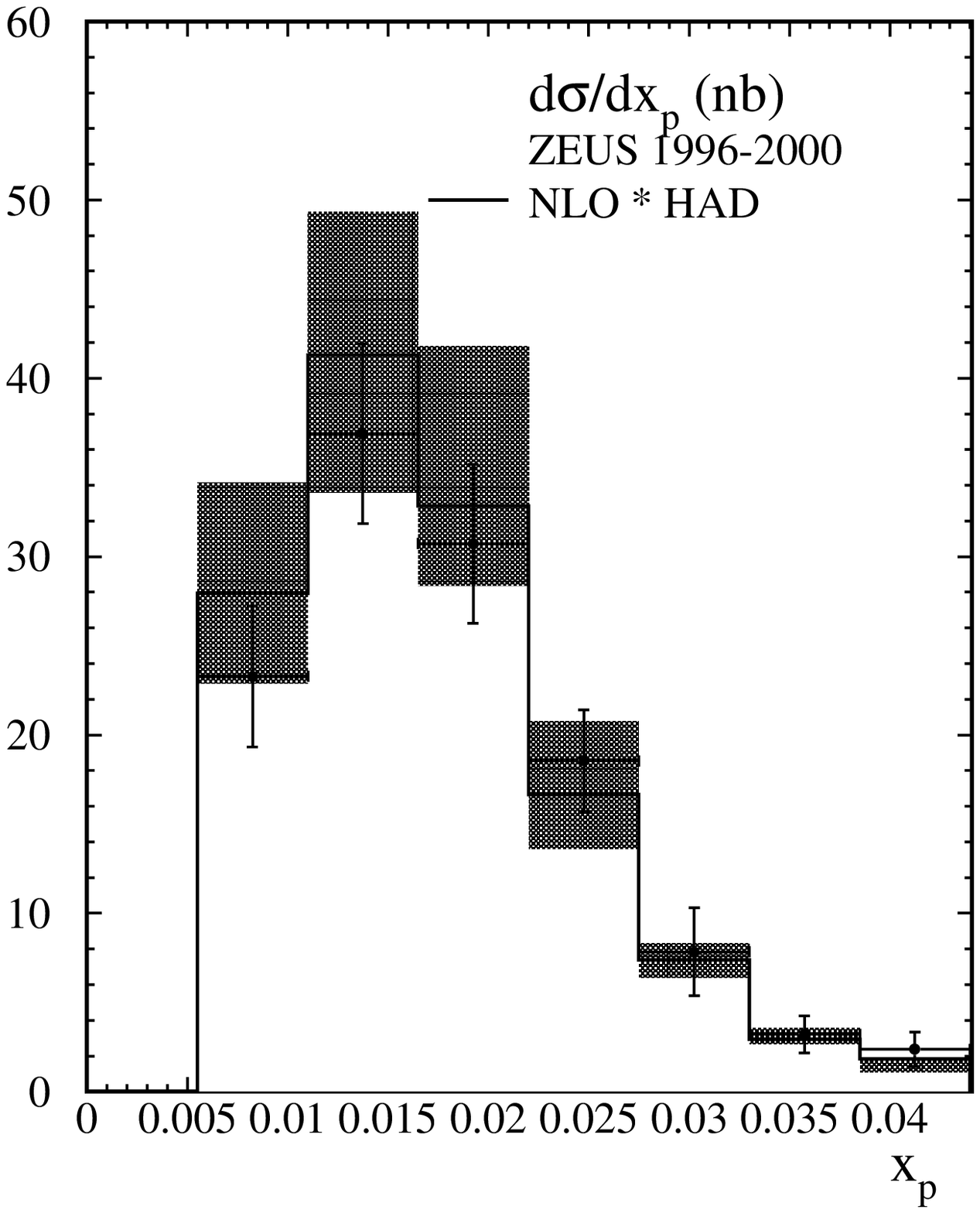,width=.55\columnwidth}
 \caption{\label{fig:3} Differential cross section $d\sigma/dx_p$ as a
function of $x_p$ compared to data of \cite{10}.}
\end{figure}
%
\begin{figure}
 \centering
\epsfig{file=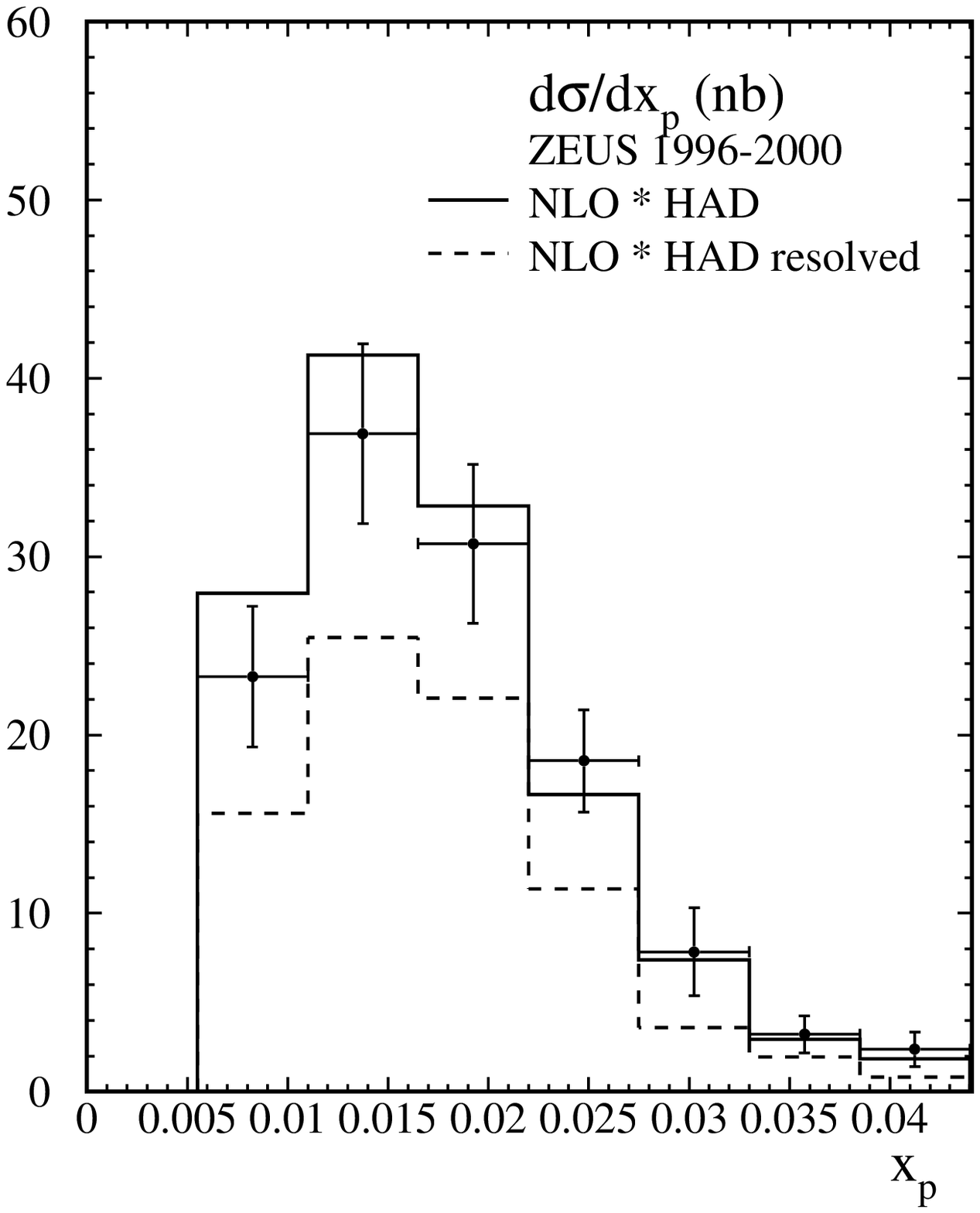,width=.55\columnwidth}
 \caption{\label{fig:4} Resolved part of the 
differential cross section $d\sigma/dx_p$ as a function of $x_p$ compared to
the full cross section and to data of \cite{10}.}
\end{figure}
In connection with inclusive dijet events obtained for our calculation with
$n_f=4$ initial quark flavours, it is of interest to know the contributions of 
charm dijet events. Therefore, we have calculated the ratio of the charm jet 
dijet cross section to the inclusive dijet cross section. This ratio is shown 
as a function of $x_p^{obs}$ (denoted as $x_p$ in the figure) in Fig. 1. The 
ratio is of order $30\%$ and almost independent of $x_p$. If the direct 
process dominated, this ratio would be 2/5 as follows from the sum of the 
squared charges of the contributing quarks. The smaller ratio in Fig. 1 is 
due to the contribution of the resolved process.
 
The differential charm dijet cross section as a function of 
$x_{\gamma}^{obs}$ (denoted $x_{\gamma}$ in the figure), which is sensitively 
dependent on the resolved process for $x_{\gamma}^{obs} < 1$, is plotted in 
Fig. 2 and compared with the data from \cite{10}. To the theoretical NLO 
predictions we have applied hadronization corrections, which have been given in
\cite{20}. In each bin the NLO cross section was multiplied by the correction 
factor $C_{had}=\sigma_{MC}^{hadron}/\sigma^{parton}_{MC}$, which is the ratio 
of the Monte Carlo (MC) cross sections after and before the hadronization 
process. In addition, we have evaluated the uncertainties in the NLO 
calculations, shown as the shaded area, originating from the variation of 
$\mu_R$ and $\mu_F$ with the parameters $\xi_R$ and $\xi_F$ 
in the range $0.5 \leq \xi_R,\xi_F \leq 2.0$ and 
$0.5 \leq \xi_R/\xi_F \leq 2.0$. The maximum (minimum) cross section is 
obtained for $\xi_R = 1.0,~\xi_F = 2.0$ ($\xi_R = 1.0,~\xi_F = 0.5$). This 
variation of scales is also used in connection with the $x_p^{obs}$ and 
$|\cos\Theta^{*}|$ distributions shown below. The low-$x_{\gamma}$ tail of the 
NLO cross section is below the data. For $x_{\gamma}^{obs} > 0.75$, the data 
are well described by the NLO prediction. In the region 
$x_{\gamma}^{obs} < 0.75$ we would have expected a better agreement with the 
data since in this region the higher order terms to the cross section 
contribute together with the resolved contribution. However, the NLO direct 
terms give a negative contribution in this region which apparently is not 
compensated enough by the resolved contribution. From this comparison we can 
conclude already that the charm part of the photon PDF as contained in the GRV 
higher order set \cite{6} is not large enough in the low $x_{\gamma}$ region. 
We shall come back to this point when we discuss other parameterizations of 
the charm PDFs of the photon.
 
The differential cross section as a function of $x_p^{obs}$ is compared in 
Fig. 3 with our NLO calculation. The NLO prediction is in good agreement with 
the data. All data points lie inside the theoretical error range, even the 
point in the largest $x_p^{obs}$ bin agrees inside the experimental error. 
Actually for the default scales $\xi_R=\xi_F=1.0$ the theoretical prediction 
agree with the data inside the experimental error. It is of interest to know 
how much of the cross section $d\sigma/dx_p^{obs}$ originates from the direct 
or resolved part of the cross section. This is shown in Fig. 4, where we have 
plotted the resolved part of $d\sigma/dx_p$ and compared with the full 
$d\sigma/dx_p$ and the experimental data of \cite{10}. We see that the resolved
part has almost the same shape and its strength is between $50$ and $60\%$ of 
the total. The direct-$\it{enriched}$ $x_p$ distribution, i.e. for 
$0.75 \leq x_{\gamma} \leq 1.0$, is between $80$ and $95\%$, depending on 
$x_p$. Thus the resolved-$\it{enriched}$ contribution of $d\sigma/dx_p$, 
i.e. for $0 \leq x_{\gamma} \leq 0.75$, is very small. Due to the lack of 
data for it in \cite{10}, a direct comparison is not possible.

%
\begin{figure}
 \centering
\epsfig{file=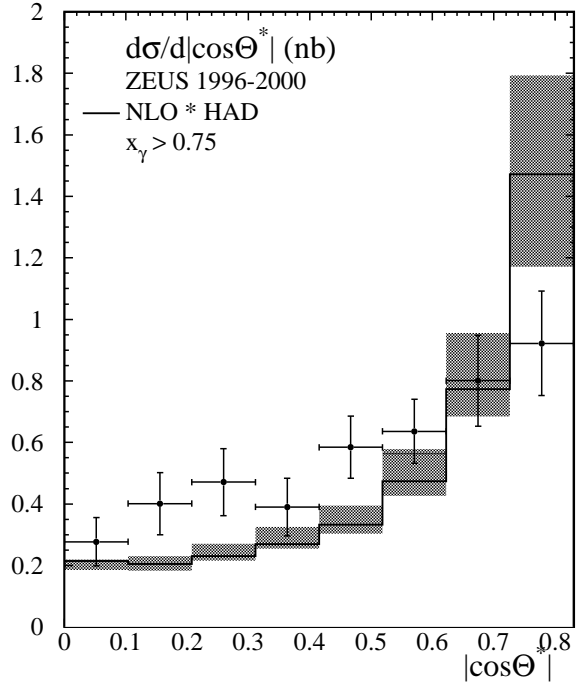,width=.55\columnwidth}
 \caption{\label{fig:5} Differential cross section $d\sigma/d|\cos\Theta^{*}|$ 
as a function of $|\cos\Theta^{*}|$ for $x_{\gamma} > 0.75$ compared to the 
data of \cite{10}.}
\end{figure}
%

%
\begin{figure}
 \centering
\epsfig{file=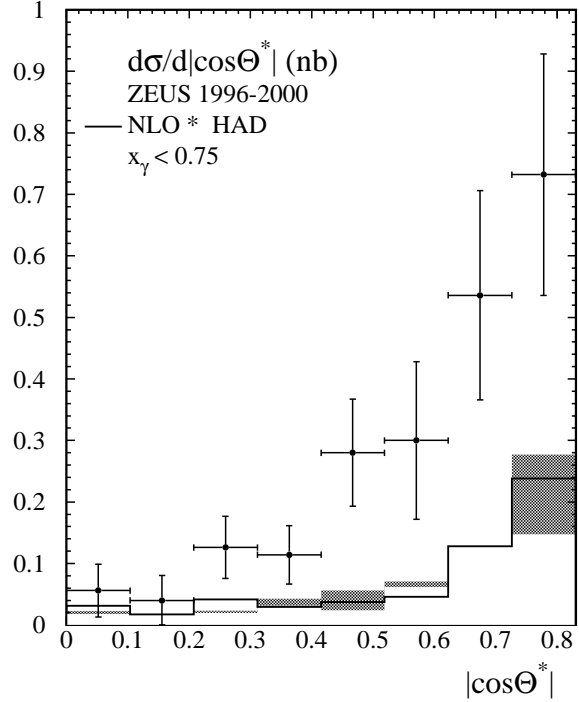,width=.55\columnwidth}
 \caption{\label{fig:6} Differential cross section $d\sigma/d|\cos\Theta^{*}|$ 
as a function of $|\cos\Theta^{*}|$ for $x_{\gamma} < 0.75$ compared to the 
data of \cite{10}.}
\end{figure}
%
In Fig. 5 and Fig. 6 we compare the NLO results for the charm dijet angular 
distribution to the ZEUS data \cite{10} as a function of $|\cos\Theta^{*}|$. 
For the high-$x_{\gamma}$ region, $0.75 < x_{\gamma} < 1.0$ (see Fig. 5), the 
NLO result is in reasonable agreement with the data, although not perfect even 
inside the theoretical error bars, which are rather small for the lower 
$|\cos\Theta^{*}|$ bins. For low $x_{\gamma}$, $0 < x_{\gamma} < 0.75$ (see 
Fig. 6), the NLO prediction is much lower than the data, except for the two 
lowest $|\cos\Theta^{*}|$ bins. This is related to the bad agreement between 
the prediction and the data for the $d\sigma/dx_{\gamma}$ cross section in the
region $x_{\gamma} < 0.75$, where apparently contributions from the resolved 
part are missing. Actually the experimental data agree much better with the LO 
prediction using the same PDFs for the proton and the photon as in the NLO 
calculation. The LO cross section for $d\sigma/d|\cos\Theta^{*}|$ is, 
depending on the $|\cos\Theta^{*}|$  bin, up to a factor between $1.1$ 
and $2.7$ larger than the NLO result.

In total we can state that our results with the GRV photon PDFs \cite{6} agree
reasonably well with the data of \cite{10} in the large-$x_{\gamma}$ region, 
but much less so in the small-$x_{\gamma}$ region, where the calculated
cross sections are too small as compared to the experimental cross sections of
\cite{10}. This is quite similar as the results obtained in \cite{10} in the
FFN scheme. From this comparison we conclude that replacing the massless charm
cross section in the calculations for the inclusive dijet photoproduction
cross section by the FFNS result would not change the result significantly.
In the FFNS calculations the resolved contribution depends only on the
photon PDFs of the light quarks and the gluon. Therefore, we do not expect that
the FFNS result will change much by choosing other photon PDFs, also in 
particular since in the FFN scheme the resolved contribution is less than in the
massless charm scheme. Therefore, a change of the theoretical result in the 
small-$x_{\gamma}$ region can be achieved only in the massless charm scheme by
changing in the resolved contribution the charm part of the photon PDFs.
To see whether this is possible we shall investigate in the follwing the 
theoretical cross sections in the small-$x_{\gamma}$ region with the more
modern photon PDFs AFG04 \cite{4} and CJK \cite{3} with the hope that the
cross section in this region will be larger.

\section{Charm densities in the photon}
\label{sec:2} 
Before we present our results, we shall take a look at the  differences between
the charm densities of the various photon PDFs as a function of the scaled 
momentum variable $x$ for the scale $Q^2=25~GeV^2$, which is the smallest 
squared scale occurring in our cross section calculations. The four charm
desities of the photon are plotted in Fig. 7 as a function of $x$ in the
interval $0 \leq  x \leq 1$. To make the comparison easier, the photon PDFs are
shown in the $DIS_{\gamma}$ scheme in order to regularize the singularity at
$x \to 1$ . Shown is the GRV charm density \cite{6}, which is already given
in the $DIS_{\gamma}$ scheme. An updated version of this is the PDF GRS 
\cite{Gluck:1999ub}, but this does not include a charm density. This is 
explicitly described by the LO FFNS contribution in $F_2^{\gamma}$, which is 
not suitable for our purpose. The next one is AFG04 \cite{4}, which is
constructed in the $\overline{MS}$ scheme and therefore transformed to the
$DIS_{\gamma}$ scheme by adding the term $C_{\gamma}$ \cite{4}. 
\begin{figure}
 \centering
 \epsfig{file=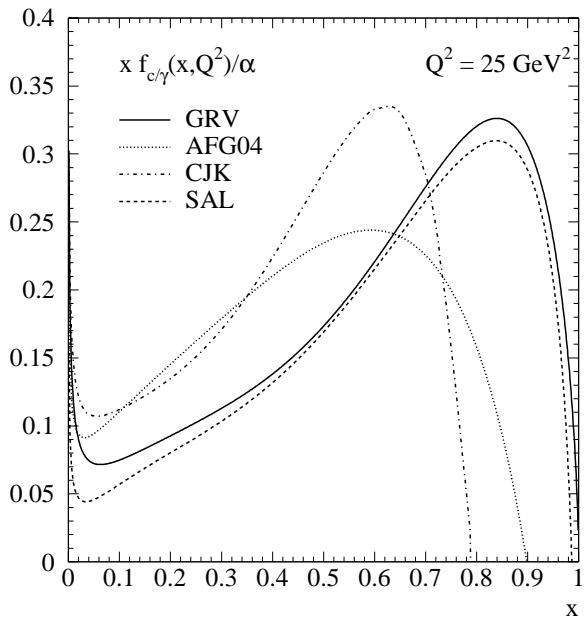,width=.55\columnwidth}
 \caption{\label{fig:7}Charm densities in the photon.}
\end{figure}
For this density $m_c=1.41$ GeV, so at $Q^2=25$ GeV$^2$, 
$x_{\rm th}=Q^2/(Q^2+4m_c^2)=0.76$. But this is not reflected in the charm PDF,
since it is constructed in the massless flavour changing scheme with $n_f=5$.
This charm density changes very little as compared to the older version AFG 
\cite{7}. It is considerably larger (smaller) than the GRV 
version at $x<0.65$ ($x>0.65$).
The CJK charm density is alo presented in the DIS$_\gamma$ scheme \cite{3}. 
Here $m_c=1.3$ GeV, so at $Q^2=25$ GeV$^2$, $x_{\rm th}=Q^2/(Q^2+4m_c^2)=0.79$.
This is clearly visible in the charm PDF, since the finite charm mass is kept.
It is considerably larger (smaller) than GRV at $x<0.7$ ($x>0.7$).
The SAL charm density is also obtained in the DIS$_\gamma$ scheme 
\cite{5}. Here $m_c=1.5$ GeV is used, but it enters only at threshold in 
the $\alpha_s$ evolution. It is very similar to GRV, but it is slightly lower 
over the full range of $x$, in particular at small $x\leq 0.05$. Because of the
similarity to our default choice GRV, we do not expect important changes
in the dijet cross sections and therefore do not consider it further.

From this comparison, we expect that the AFG04 and even more so the CJK
charm photon PDFs should yield larger cross sections than the GRV version in
the small-$x_{\gamma}^{obs}$ region and somewhat smaller coss sections in the
large-$x_{\gamma}^{obs}$ region. How the cross sections for these three
photon PDFs, GRV, AFG04 and CJK, compare with each other and with the ZEUS 
data will be shown in the next section. 
%
\begin{figure}
\centering
\epsfig{file=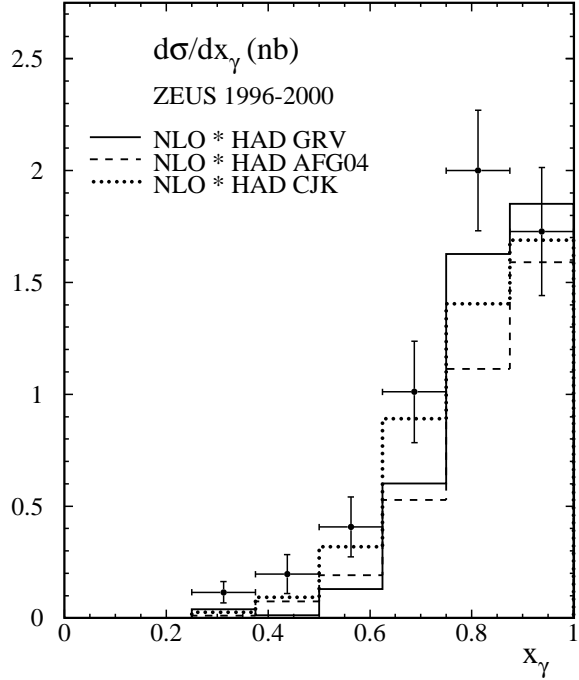,width=.55\columnwidth}
\caption{\label{fig:8} Differential cross sections $\d\sigma/dx_{\gamma}$
for GRV, AFG04 and CJK photon PDFs compared to data of \cite{10}.}
\end{figure}
%
\section{Results for GRV, AFG04 and CJK Photon PDFs}
\label{sec:5}
First we show the cross section $d\sigma/dx_{\gamma}$ in Fig. 8 for the three
photon PDFs GRV, AFG04 and CJK, where we expect better agreement with the
experimental cross section for $x_{\gamma} < 0.75$  with CJK. This is indeed
the case. Whereas in the two largest $x_{\gamma}$-bins there is little change 
between GRV and CJK, the rest of the $x_{\gamma}$-bins have larger cross 
sections for CJK and agree now much better with the ZEUS data. 
For completeness we also show the equivalent comparison for the cross section
$d\sigma/dx_p$ (Fig. 9). Here all three photon PDFs yield almost the same
cross section and agree equally well with the data.
%
\begin{figure}
\centering
\epsfig{file=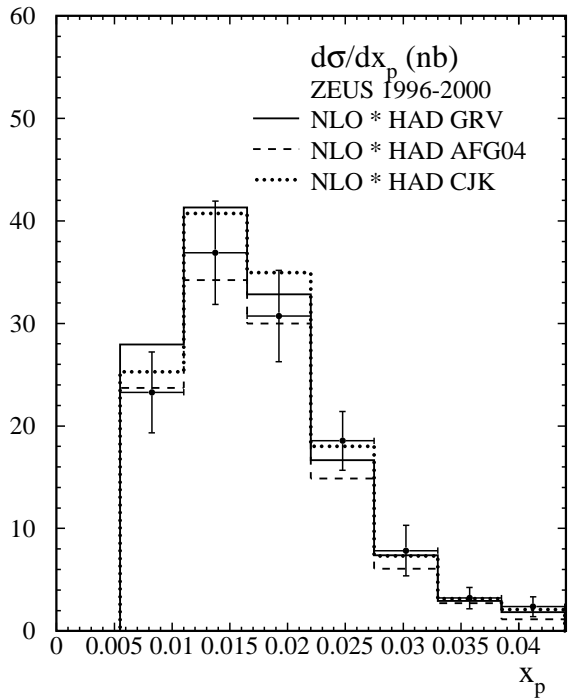,width=.55\columnwidth}
\caption{\label{fig:9} Differential cross section $d\sigma/dx_p$ for GRV, AFG04
and CJK photon PDFs compared to data of \cite{10}.}
\end{figure}
%
A similar pattern occurs for the $|\cos\Theta^{*}|$ distribution in the large
$x_{\gamma}$ region ($x_{\gamma} > 0.75$). The cross section for the three
PDFs are very similar and agree equally with the ZEUS data. Only at
$|\cos\Theta^{*}| \sim 0.8$ the cross setion for AFG04 and CJK is somewhat
reduced (see Fig. 10).
%
\begin{figure}
\centering
\epsfig{file=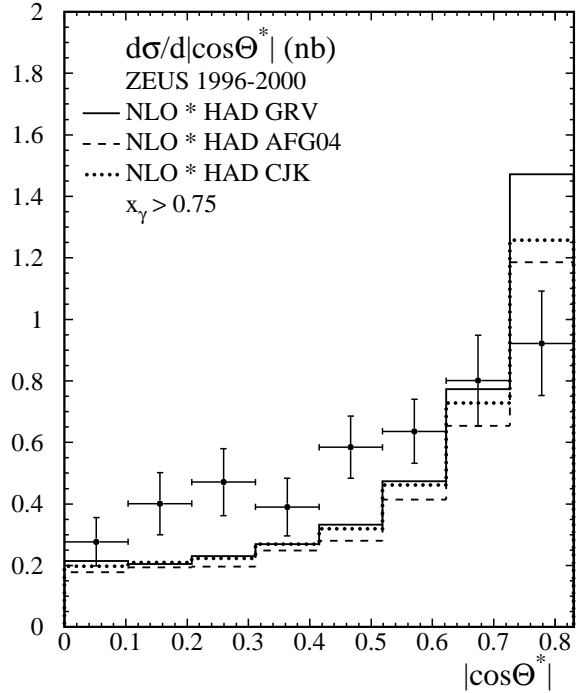,width=.55\columnwidth}
\caption{\label{fig:10} Differential cross section $d\sigma/d|\cos\Theta^{*}|$
with the constraint $x_{\gamma} > 0.75$
for GRV, AFG04 and CJK photon PDFs compared to data of \cite{10}.}
\end{figure}
The critical cross section is $d\sigma/d|\cos\Theta^{*}|$ for 
$x_{\gamma} < 0.75$ shown in Fig. 11. Here the agreement with the data was
bad for GRV, in particular for $|\cos\Theta^{*}| > 0.3$. In this region
the data agree now better with the CJK prediction as expected, although not
perfectly. The cross section for CJK is more than a factor of two
larger than for GRV in this region of $|\cos\Theta^{*}|$. The result for
AFG04 is very similar to that of GRV and does not lead to any improvement
compared to the experimental results.
%
\begin{figure}
\centering
\epsfig{file=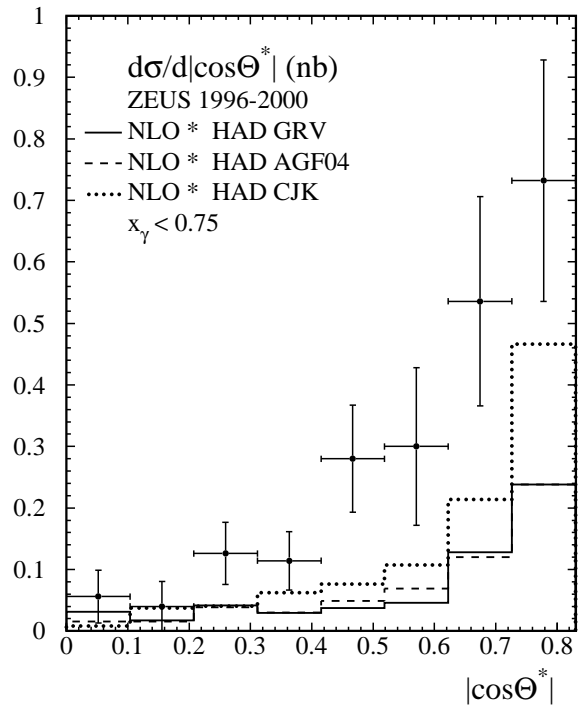,width=.55\columnwidth}
\caption{\label{fig:11} Differential cross section $d\sigma/d|\cos\Theta^{*}|$
with the constraint $x_{\gamma} < 0.75$
for GRV, AFG04 and CJK photon PDFs compared to data of \cite{10}.}
\end{figure}
%

%
%

%
%

As mentioned in the introduction, the dijet photoproduction cross section as a
function of $x_\gamma$ has been measured by the ZEUS collaboration also for all
flavours \cite{8}, and the comparison with NLO calculations showed in general good
agreement for all tested photon PDFs. However, the NLO predictions obtained with
the CJK photon PDFs were larger than the data at low $x_\gamma\leq 0.6$ and
$E_T\simeq20$ GeV, while they agreed better with the data than those obtained
with the other PDFs in the second and larger $E_T$-bins. It may thus be necessary
to compensate the higher charm-quark density favoured by the charm dijet analysis
with a smaller up-quark density at lower values $x_\gamma$ (cf.\ also Fig.\ 14
in \cite{3}). Here one must keep in mind that the full dijet analysis has been
performed at larger values of $E_T\geq 20$ GeV (or $Q$) than the charm dijet
analysis, where $E_T\geq5$ GeV (see also Fig.\ 15 in \cite{3} for the evolution
of the CJK charm density with $Q^2$).

\section{Conclusion}
\label{sec:6}

In summary, we have demonstrated that the contribution of charm quarks to
photoproduced dijets at HERA is substantial with charm quarks (or mesons)
accounting for about one-third of all photoproduced dijets. This is in good
agreement with the naive estimate of 2/5 from the sum of the squared quark
charges coupling to the photon and offers the possibility to constrain the
charm quark density in the photon.

To this end, we have computed the charm dijet photoproduction cross section at
NLO of QCD in the zero-mass variable flavour number scheme, i.e.\ with active
charm quarks in the proton and photon. This approach is justified by the fact
that at large values of transverse energy the charm quark mass may safely be
neglected, so that the contributions of collinear charm quark excitations may be
resummed into parton densities in the photon and proton.

Our theoretical results were compared to recent measurements from the ZEUS
experiment at HERA. The distributions in the photon and proton
momentum fractions and the dijet scattering angle agreed well with the data, in
particular for large momentum fractions of the partons in the photon, where
direct photon processes dominate. At low momentum fractions, the predictions were
quite sensitive to the charm content in the photon.

We demonstrated that the experimental data favoured parameterizations, like the
one by the CJK collaboration, with a substantial charm quark density. Since the
total dijet cross section is overestimated by this parameterization, at least at
low momentum fractions of the partons in the photon, it may, however, be necessary
to compensate the higher charm-quark density with a smaller up-quark density.
The logical next step would thus be to perform a global analysis of
$F_2^\gamma(x)$ and dijet photoproduction data to better constrain the different
flavours in the photon PDFs.

%

\end{document}